\documentclass{aa}  
\usepackage{array}
\usepackage{textcomp, gensymb}
\usepackage{comment}
\usepackage{xcolor}
\usepackage{footmisc}
\usepackage{tablefootnote}
\usepackage{float}
\usepackage{commath}
\usepackage{sidecap}

\usepackage{calrsfs}
\DeclareMathAlphabet{\pazocal}{OMS}{zplm}{m}{n}

\usepackage{graphicx}
\usepackage{txfonts}
\usepackage{hyperref}
\hypersetup{colorlinks,linkcolor={blue},citecolor={blue},urlcolor={blue}}
\begin{document} 

   \title{Magnetically arrested disk flux eruption events to describe \\ SgrA* flares} 
    \titlerunning{MAD flux eruption events to describe SgrA* flares}
   \author{Eleni Antonopoulou\inst{1}\fnmsep\inst{2}, Argyrios Loules\inst{1}, Antonios Nathanail\inst{2}}
   \institute{Department of Physics, National and Kapodistrian University of Athens, University Campus, GR 15784 Zografos, Greece
    \and
    Research Center for Astronomy and Applied Mathematics, Academy of Athens, Soranou Efesiou 4, 115 27 Athens, Greece
             }
   \date{Received December 16, 2024; accepted February 5, 2025}
  \abstract
   {Magnetically arrested disks (MADs) are among the most suitable candidates for describing the gas accretion and observed emission in the vicinity of supermassive black holes.} 
   {We aim to establish a direct correlation between the quasiperiodic flux eruption events, characteristic of MAD accretion disk simulations, and the observed flaring behavior in the Galactic center.}
   {We employed a MAD accretion disk with a distinct counterclockwise rotation and investigated the evolution of magnetized flux tubes generated during a prominent flux eruption event. Although these flux tubes demonstrate a clockwise pattern, they experience significant drag from the accretion disk's rotation. We modeled the motion of hot spots, formed on the disk's equatorial plane due to magnetic reconnection, as they travel along the magnetized flux tubes at a fraction of the speed of light.}
    {Hot spots with a relativistic ejection velocity are able to balance out the counterclockwise drag of the flux tube's foot-point on the disk and demonstrate a clockwise motion in the sky, which is in good agreement with the near-infrared flares in the Galactic center. In addition, our flare models favor face-on inclinations in the ranges $[0\degree, 34\degree]$ and $[163\degree, 180\degree]$ for SgrA*. }
  {The flux eruption events that arise naturally in the MAD accretion state provide a promising framework for reproducing the observed flaring behavior in the vicinity of SgrA*.}
   \maketitle
%
\section{Introduction}
Sagittarius A* (SgrA*), the supermassive black hole in the Galactic center, experiences intense flaring events visible in the near-infrared (NIR) wavelengths \citep{genzel2003,Ghez_2004} and X-ray \citep{baganoff2001}  several times a day. 
Recent observations of the GRAVITY instrument at the Very Large Telescope have shed light on both the astrometry and the polarimetric signature of bright NIR flares \citep{GRAVITY_2018,GRAVITY_2020,GRAVITY_2023}.
The most prominent events exhibit a relativistic clockwise motion a few gravitational radii from the supermassive black hole, accompanied by a matching evolution in polarization angle over a period of approximately one hour. 
These profound observations allow significant constraints to be placed on the orbital motion of hot spots in the vicinity of SgrA* \citep{GRAVITY_2020} and suggest the existence of strong poloidal magnetic fields in the emitting region of our Galactic center \citep{GRAVITY_2020_polarimetry}.
\par
The origin of Galactic center hot spots was first investigated by \cite{Matsumoto_2020}, who point out the super-Keplerian pattern speed of the underlying accretion flow.
Since then,
several models have been employed to investigate the dynamics of the highly polarized NIR emission and reproduce the observed flaring behavior in the vicinity of SgrA* \citep{Tursunov_2020,Ball_2021,Vos_2022,vincent_2023,yfantis_2023,Lin_2023,Aimar_2023,Huang_2024,kocherlakota_2024,Lin_2024,Antonopoulou_2024}.
The aforementioned studies have provided significant insights into the parametric properties of such flares. These studies favor super-Keplerian motion and nearly face-on observation angles, examine the impact of considering ejected hot spot configurations and electromagnetic interaction within the inner magnetosphere, and highlight the nonthermal properties of the emitting electron population.
Nevertheless, the precise mechanism governing the ejection and energetic emission of hot spots in the Galactic center is still unknown.
\par
Magnetically arrested disk (MAD) models are among the most suitable candidates for describing active galactic nuclei that exhibit strong jets \citep{Bisnovatyi_Ruzmaikin_1974,Narayan_2003,Sukova_2021,CruzOsorio_2022,Fromm_2022}.  
In addition, large-scale simulations of stellar mass feeding in the Galactic center have also successfully reproduced MAD disks \citep{Ressler_2019}.
In the MAD state, poloidal magnetic flux is advected toward the black hole by the accreting gas until the magnetic pressure becomes strong enough to balance the ram pressure of the disk \citep{Igumenshchev_2003,Igumenshchev_2008,Tchekhovskoy_2011}.
When the two competing forces reach equipartition, gas accretion is significantly hindered and can only occur via three-dimensional non-axisymmetric  processes, such as the magnetic Rayleigh-Taylor instability \citep{Papadopoulos_2018}.
\par 
General-relativistic magnetohydrodynamic (GRMHD) simulations of MAD accretion disks reveal a quasiperiodic cycle of magnetic flux accumulation and ejection, a process we define as a flux eruption event. 
In each flux eruption event, magnetized bundles or hot spots are expelled into the disk.
\begin{figure*}[h]
\centering
\includegraphics[width=0.75\textwidth]{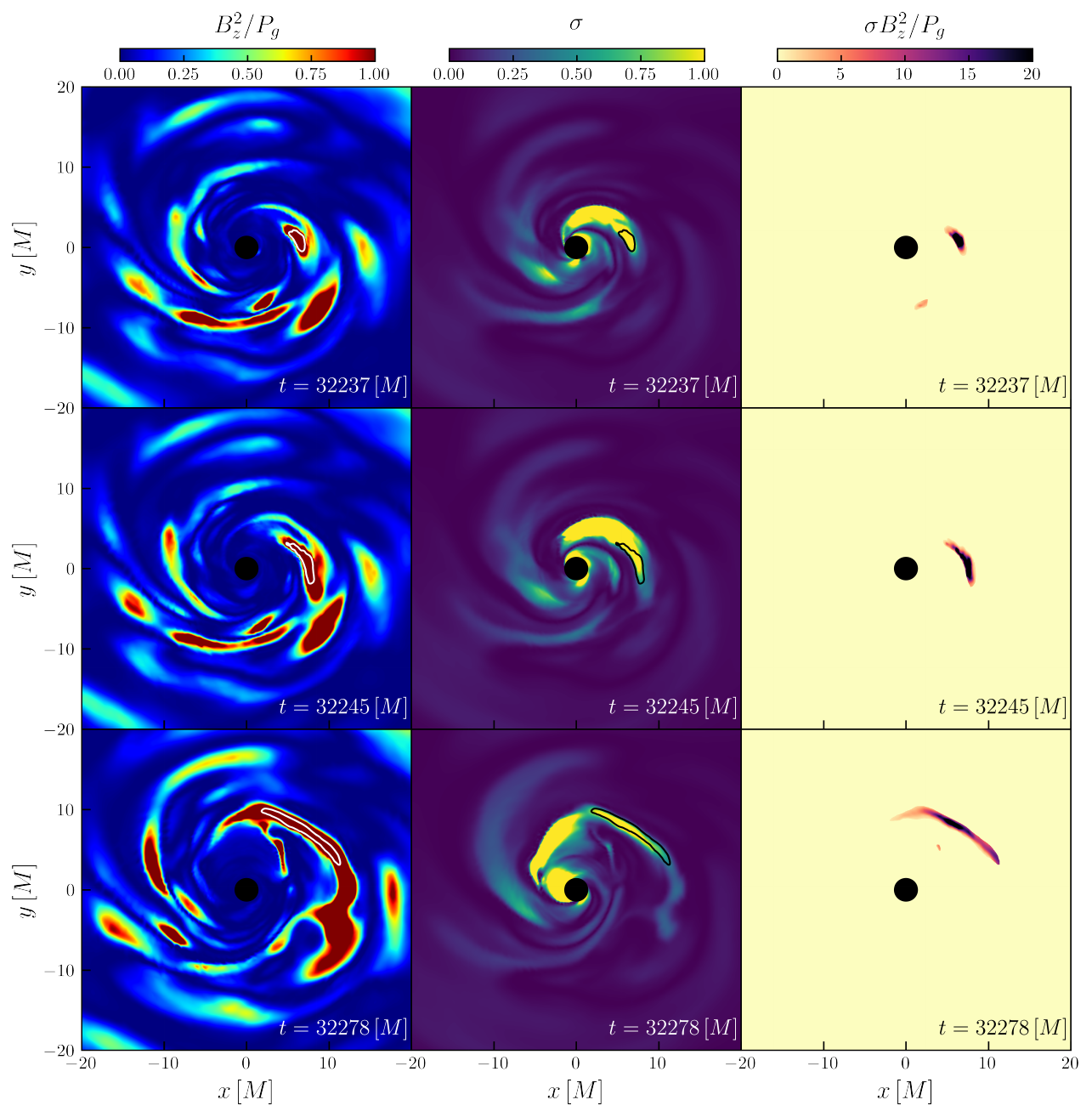}
\caption{Evolution of quantities on the equatorial plane throughout the selected flux eruption event. The first column illustrates the ratio of vertical magnetic field strength to gas pressure, the second column the magnetization, and the third column their product, $\sigma B_z^2/P_g$, used for the identification of magnetic reconnection sites on the equatorial plane. The white and black contours in the first two columns outline the magnetic reconnection sites illustrated in the third column.}
\label{Fig:Flux_eruption}
\end{figure*}
One of the main energy-dissipation mechanisms responsible for hot spot generation close to the event horizon of an accreting black hole is magnetic reconnection \citep{Ripperda_2022,Nathanail_2022}.
Turbulence and instabilities in the accretion flows of MAD models generate large current sheets with alternating magnetic field polarity, which are highly efficient sites for particle acceleration \citep{Kagan_2015}. 
Magnetic reconnection in these sheets creates chains of magnetized hot spots, each a few gravitational radii in size \citep{Loureiro_2007, Fermo_2010, Uzdensky_2010, Huang_2012, Loureiro_2012, Takamoto_2013, Nathanail_2020, Chatterjee_2021,Sukova_2021}. 
While these hot spots have a minimal impact on the overall accretion flow variability, they are linked to observed variability in NIR and X-ray wavelengths and episodic flares near our Galactic center \citep{Moscibr_2007,Dimitropoulos_2024}. \par
Consequently, MAD models are well suited for flux tube formation and energetic flare emission, and provide promising candidates for reproducing the observed flares in the vicinity of SgrA*.
The magnetic reconnection events that naturally arise in the MAD state have been shown to produce episodic flares that demonstrate the continuous rotation and typical timescales characteristic of NIR flares in the Galactic center \citep{Dexter_2020}, as well as the polarization signature and timescales of flares at submillimeter wavelengths \citep{mahdi_2023}.
Although the magnetic energy associated with such events is capable of powering the observed IR and X-ray flares, the motion of the generated flux tubes within the accretion flow is substantially sub-Keplerian, in contrast with the GRAVITY observations \citep{Porth_2021}. 
\cite{Lin_2024} identified flux ropes consistent with the theoretical "coronal mass ejection" model \citep{Yuan_2009} within GRMHD simulations of black hole accretion disks and semi-analytically reconstructed their evolved trajectories to fit the July 2018 flare.
\par
This study extends previous research on MAD models by conducting a three-dimensional GRMHD simulation focused on isolating flux tubes from energetic flux eruption events and modeling the ejected hot spot emission, which is in accordance with the most accurate astrophysical models for SgrA*. 
Although \cite{EHTV2022} were not able to identify a preferred position angle for the source, they determined a cluster of strongly magnetized (MAD) models with positive spin and low inclination that successfully pass all but two of the observational constraints for SgrA* across a broad frequency range. 
We employed a highly spinning Kerr black hole with parameter $a=0.94$ and a counterclockwise rotation (on the plane of the sky) for the magnetized accretion disk, in agreement with one of the most promising MAD models.
In this configuration, the generated flux bundles experience significant bending due to rotation and have a distinct clockwise pattern during the flux eruption events.
Our analysis emphasizes the role of MAD configurations in modeling gas accretion near SgrA* and directly links simulated hot spot ejection to the observed flaring activity in our Galactic center. 
Our findings allow us to set strict constraints on disk dynamics, and reveal that a counterclockwise rotation of the disk plasma, coupled with the clockwise motion of hot spots atop the flux bundle, is capable of replicating the observed NIR flares.
\par
This paper is organized as follows: 
Section \ref{Sec:Methodology} outlines the methodology for identifying promising flux eruption events and modeling flares in MAD simulations (details on the modeling process can be found in Appendix \ref{Append:Modeling_MAD_flares}).
Section \ref{Sec:Results} compares the flare models to the observations of \cite{GRAVITY_2018} and sets constraints on the inclination of SgrA* (a detailed discussion is provided in Appendix \ref{Append:inclination_SgrA}).
Section \ref{Sec:Conclusions} contains the conclusions of our analysis.
The numerical setup for the GRMHD simulation and general-relativistic  radiative transfer calculations is described in Appendix \ref{Append:Numerical_Setup}, whereas the time series of the mass accretion rate and normalized magnetic flux are given in Appendix \ref{Append:MAD_Time_Series}. 
Appendix \ref{Append:Flare_Energetics} estimates the flare energetics, and Appendix \ref{Append:GRAVITY_first_observation} focuses on replicating the first observational data points. 
In Appendix \ref{Append:clockwise_flares}, we briefly investigate the scenario of clockwise disk rotation.
\section{Methodology} \label{Sec:Methodology}
We investigated the evolution of a three-dimensional MAD accretion disk simulation, focusing on the generation and development of energetic flux tubes during flux eruption events. 
These events are initiated by intense magnetic reconnection on the disk's equatorial plane, producing bundles of energized particles.
We assumed that the produced bundles travel along the magnetized flux tubes. 
Flux tubes in this context are typically dominated by coherent, large-scale vertical magnetic fields, resulting in significant magnetic field strength. 
Their foot-points on the disk are also highly magnetized\footnote{$\sigma = B_{co}^2/\rho$, where $B_{co}$ is the magnetic field in the comoving frame and $\rho$ is the density of the accretion disk.}. 
Notably, magnetic flux from previous flux eruption events remains within the disk, continuing to evolve (move around) and potentially 
influence subsequent eruptions, creating regions with "stagnating" magnetic flux.
Therefore, we initially shifted our focus to the accretion disk's equator and searched for regions with a high value of magnetization, $\sigma \geq 0.4$, and a large ratio of vertical magnetic field strength over thermal pressure, $B_z^2/P_g \geq 0.8$, that have been produced by newly generated flares.
\par
\begin{figure}[t]
\centering
\includegraphics[width=0.45\textwidth]{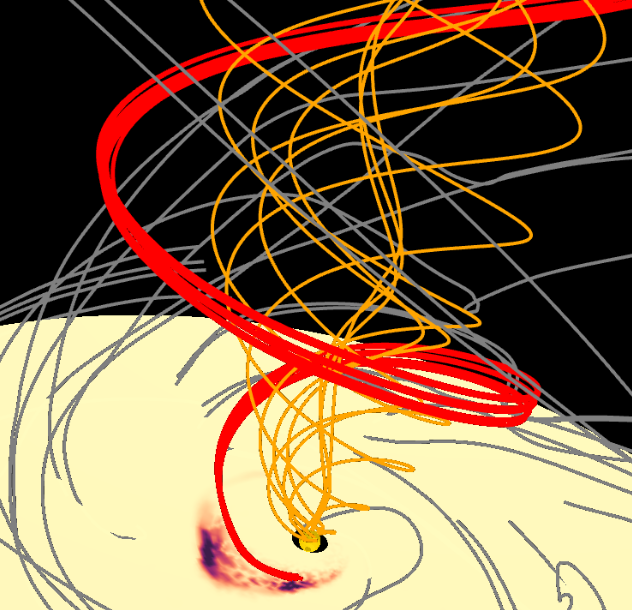} 
\caption{Three-dimensional representation of the accretion flow and magnetic field topology during a flux eruption event. The color in the equatorial plane corresponds to the product of the magnetization times the vertical magnetic field strength, divided by the gas pressure ($\sigma\times B_z^2/P_g$), highlighting newly generated flux bundles (red lines) created by the flux eruption event. The red field lines are drawn from the peak of the quantity plotted in color, corresponding to the most energetic part of the flux bundle. The gray lines represent the magnetic field lines of the accretion disk. The orange lines show the magnetized funnel field lines, which originates from the black hole and had been connected to the flux bundle before the event.}
\label{Fig:3D_GRMHD}
\end{figure}
Our analysis centers on a specific flux eruption event occurring at 
$32228\,M$ and originating approximately at $4.7\,r_g$ from the supermassive black hole. 
Figure \ref{Fig:Flux_eruption} illustrates this event by displaying the ratio of vertical magnetic field strength over gas pressure (first column), the magnetization (second column), and their product (third column) evaluated on the equatorial plane of the accretion disk during the selected flux eruption event.
Initially confined, the active region quickly expands in both azimuthal directions.
As shown in Fig. \ref{Fig:Flux_eruption}, this region exhibits both an orbital motion due to the accretion disk's rotation and an outward motion toward larger orbital radii, remaining active for over $90\,M$ \citep{Porth_2021}. 
This activity makes it an ideal candidate to model the observed flaring behavior in our Galactic center.
In addition to this eruption, other flux eruption events appear throughout the simulation. These events tend to stagnate within the disk, moving passively with the flow. They can be identified by their strong vertical magnetic field and characteristic peaks in $B_z^2/P_g$ (first column of Fig. \ref{Fig:Flux_eruption}). 
Over time, their magnetization decreases, indicating that only recently generated flux tubes from fresh eruptions maintain high magnetization (second column of Fig. \ref{Fig:Flux_eruption}). 
The product of these quantities, shown in the third column of Fig. \ref{Fig:Flux_eruption}, marks where newly generated flux tubes are introduced into the disk via a flux eruption event.
\par
\begin{figure*}
\sidecaption
\includegraphics[width=5.75cm]{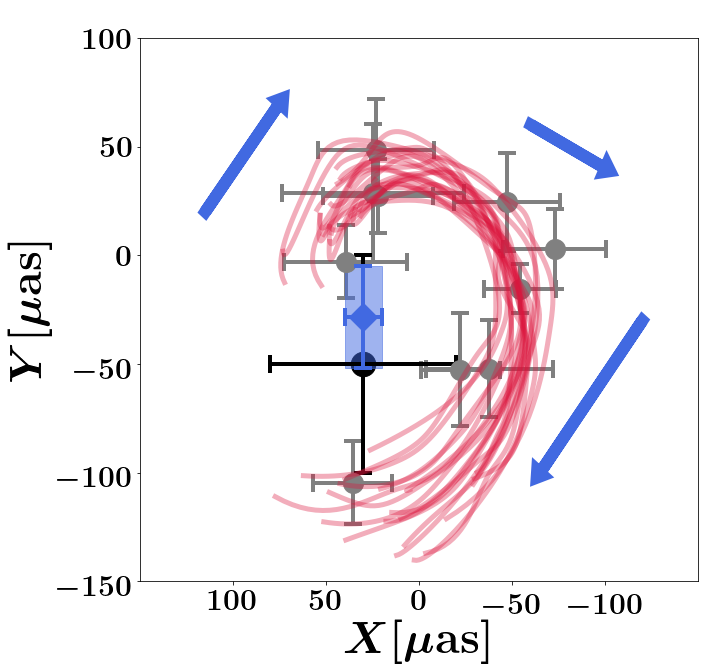}
\hspace{0.0cm}
\includegraphics[width=6.65cm]{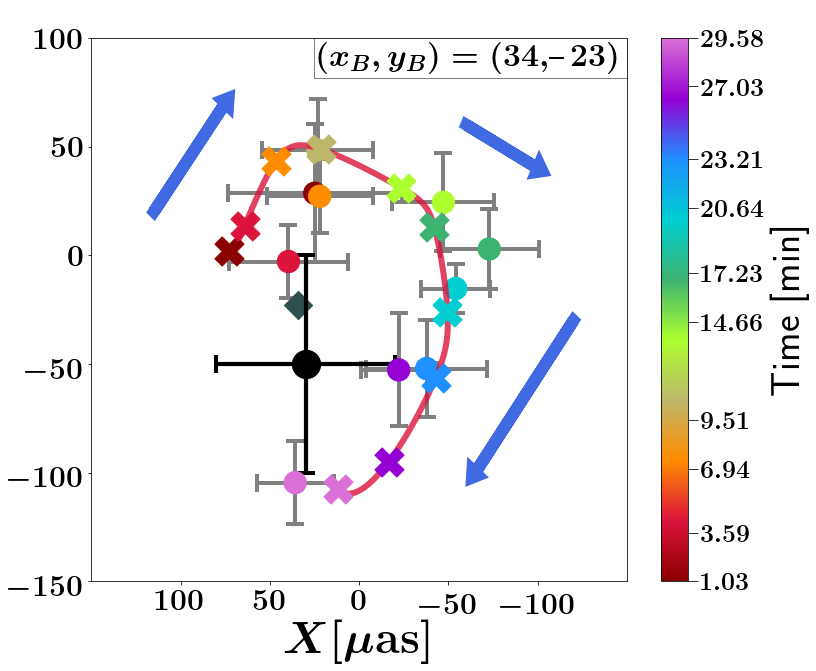} 
\caption{Hot spot trajectories along the magnetized flux tubes generated during the prominent flaring event overlapped with the observations of July 22, 2018 (gray and colored circles, respectively). The direction of motion is indicated by blue arrows. The position of SgrA* in the sky is denoted by a black cross. \textit{Left panel}: Flare models for hot spot ejection velocities between $0.5c$ and $0.8c$ and foot-point orbital velocities ranging from $0.1u_K-0.4u_K$. The best-fit black hole position, which is slightly different for each model, is illustrated by a blue rectangle. \textit{Right panel}: Flare model that provides the best fit to the GRAVITY observations. The position of the black hole is depicted by a black diamond.}
\label{Fig:Flare_Observations}
\end{figure*}
After tracking the evolution of the magnetic reconnection site on the equatorial plane, we examined the three-dimensional morphology of the energetic flux tubes generated during the selected flux eruption event. 
Figure \ref{Fig:3D_GRMHD} presents a three-dimensional view of the magnetic field lines, showing in red the newly generated flux bundles from the flux eruption event initially identified on the equatorial plane. 
The disk’s magnetic field lines are shown in gray, while the magnetized funnel field lines — connected to the red flux bundle prior to the eruption and originating from the black hole's event horizon — are highlighted in orange.
This visualization captures the shape and dynamics of the magnetized flux tube originating from the active region identified in the third panel of Fig. \ref{Fig:Flux_eruption}, displaying its evolution at several points throughout the prominent flux eruption event.
This approach allows us to track the morphology of magnetic field lines within the three-dimensional space-time of the MAD accretion disk, providing the framework for analyzing the evolution of the hot spot trajectory. 
\par
Notably, each hot spot model corresponds to a fixed flux tube structure captured at a specific timestamp in the MAD simulation, while the overall evolution of the flux tube's shape throughout the flux eruption event is represented by different models (see the left panel of Fig. \ref{Fig:Flare_Observations}).
This study considers spherical hot spots with an effective emission radius of $1M$ that are generated at the equatorial plane of the disk and propagate along the magnetized flux tube with a constant ejection velocity of a fraction of the speed of light. 
In addition, we tracked the position of the flux tube's foot-point on the accretion disk during the selected flux eruption event and implemented both the outward motion and sub-Keplerian orbital rotation in the flare models, assuming solid-body rotation for the flux bundle.
We note that other models, which reconstruct trajectories from the toroidal velocity from a numerical simulation, suggest that the hot spot moves in the same direction as the rotation of the flow \citep{Nathanail_2020, Lin_2024}. In contrast, our model predicts that the disk plasma and the hot spot, which is anchored to the flux bundle, move in the opposite direction.
More details on the modeling procedure can be found in Appendix \ref{Append:Modeling_MAD_flares}.
\section{Results} \label{Sec:Results}
%
%
According to the analysis outlined in Sect. \ref{Sec:Methodology}, the prominent flux eruption event illustrated in Fig. \ref{Fig:Flux_eruption} constitutes a promising candidate for replicating the observed flaring behavior in the vicinity of SgrA*.
We tracked the trajectory of the hot spot along the magnetized flux tubes generated during the selected flux eruption event and calculated the emitted radiation for individual models.
Unless stated otherwise, we employed a face-on observer inclination in our calculations.
We note that the focal point of this study is the most investigated flare of July 22, 2018. 
The observed flare positions represent the average of the "Waisberg" and "Pfuhl" analyses, thus incorporating the individual differences between the \cite{GRAVITY_2018} reduction methodologies (a more detailed discussion is given in Appendix \ref{Append:GRAVITY_first_observation}).
\par
The left panel of Fig. \ref{Fig:Flare_Observations} illustrates the flaring behavior for various hot spot models $-$and therefore several flux tube configurations throughout the selected flux eruption event$-$ overlapped with the observations of July 22, 2018.
Each flare model corresponds to a different ejection for the hot spot, ranging from $0.5c$ to $0.8c$, and orbital velocity for the flux tube's foot-point on the disk, between $0.1u_K$ and $0.4u_K$ based on the simulation (for more details see Appendix \ref{Append:Modeling_MAD_flares}).
Even though the MAD accretion disk possesses a distinct counterclockwise rotation, the observed hot spot trajectories demonstrate a clockwise motion in the sky, due to the relativistic velocity of the ejected hot spots.
In particular, the flares modeled throughout the flux eruption event demonstrate a clear trend, whose general characteristics are in good agreement with the observed flaring behavior in our Galactic center.
\par
The right panel of Fig. \ref{Fig:Flare_Observations} depicts a flare model yielding very good agreement with the observations of July 22, 2018.
In this particular model, the ejected hot spot traverses along the magnetized flux tube with a relativistic velocity of $0.8c$, whereas the flux tube's foot-point on the accretion disk is rotating with an orbital velocity equal to $0.1u_{K}$.
The illustrated trajectory demonstrates a good correlation with the general trend of the GRAVITY observations and falls within the associated error bars for all but one of the observed flaring positions (see the purple point at $t=27.03min$).
Although this particular flare model does not capture the kinematics of the first data points (dark red point at $t=1.03$ min), it still falls within the corresponding error bars, since the uncertainty for the first flaring position is significantly big.
It is important to note, however, that a number of models illustrated in Fig. \ref{Fig:Flare_Observations} are capable of reproducing the first data points of the July 22 flare.
A more detailed discussion on the topic is given in Appendix \ref{Append:GRAVITY_first_observation}.
\par
The final step of our analysis was to create an image of the best-fit flare model (see the right panel of Fig. \ref{Fig:Flare_Observations}) for several observer inclinations.
We considered an evenly spaced range of observation angles up to 85°, along with their supplementary angles. 
At edge-on inclinations, the orbits exhibit significant deformation, either shrinking or stretching in the plane of the sky, which pushes the hot spot trajectories outside the bounds of the observed flaring emission (see Fig. \ref{Fig:Inclinations}). 
Our analysis highlights the strong preference the GRAVITY observations demonstrate for face-on inclinations and constrains the inclination of SgrA* to the range between $[0\degree, 34\degree]$ and $[163\degree, 180\degree]$.
A more detailed discussion on the effect of inclination on the resulting flare emission is given in Appendix \ref{Append:inclination_SgrA}.
\section{Conclusions and discussion} \label{Sec:Conclusions}
This study explored the dynamics of flux eruption events in MAD accretion flows, with a particular focus on the formation, evolution, and motion of hot spots generated from magnetic reconnection during a flux eruption event. 
We conducted a three-dimensional GRMHD simulation $-$ motivated by the best-bet models of \cite{EHTV2022} $-$ in order to identify energetic flux tubes and track their morphology and evolution within the disk’s highly magnetized environment. 
Our calculations take into account both the orbital motion of the flux tube's foot-point on the disk and the outward radial motion identified in the simulation while maintaining a constant ejection velocity for the hot spots. 
The focal point of this study is the incorporation of a counterclockwise motion for the MAD accretion disk, in contrast to the clockwise NIR flares observed with GRAVITY.
This distinct direction of rotation is consistent with the assumption that the rotation measure can be entirely attributed to an internal Faraday rotation in the polarimetric analysis of \cite{EHTVIII2024}.
\par
By modeling specific flux eruption events, we set constraints on disk plasma rotation and examined the role of magnetic field geometry in shaping the flare trajectories.
The flux eruption events observed in our simulation closely align with flaring activity seen near SgrA*, providing a promising framework for understanding high-energy phenomena in the vicinity of supermassive black holes. 
This study highlights the distinct counterclockwise flow of the flux bundles and the clockwise motion of the energetic particles within them, driven by the rotational dynamics of the disk, as well as the bending of the magnetized tube.
Our findings further highlight that MAD configurations offer a robust description of the variable environment of SgrA*, while at the same time providing a direct link between the multiwavelength observations of the Galactic center and the episodic flares observed in the vicinity of the supermassive black hole. 
\par
However, to differentiate this model from other competing scenarios, specific "smoking gun" observations are required. Key distinguishing features would include the absence of a steady-state jet and the presence of an intermittent flow transitioning between quiescence and flaring activity. Crucially, observations at 43 and 86 ${\rm GHz}$  \citep{Nathanail_2022b} could reveal a significant signature of these behaviors, particularly through the detection or non-detection of a structure near the black hole's base from a well-defined jet. A structure different from what has been observed at the base of a jet, for example in M87, would provide direct evidence of the unique magnetic field geometry proposed.
\par
The quasiperiodic flux eruption events $-$ characteristic of MAD simulations $-$ and the flare model presented in this work offer valuable insights into the energetic phenomena near supermassive black holes beyond the Galactic center.
Consequently, an interesting application of our flare model would be the quasiperiodic outflows observed in extragalactic nuclei.
\cite{Pasham_2024} report the discovery of an ultra-fast quasiperiodic outflow from a low-luminosity active galactic nucleus, which they attributed to a binary black hole system.
In particular, they suggest that the intermediate-mass black hole secondary is orbiting the supermassive black hole primary and expelling blobs of gas into the magnetized funnel via a mechanism closely related to that in our hot spot ejection model.
\par
Last but not least, our study places strict constraints on the inclination of SgrA*, which lies in the face-on range between $[0\degree, 34\degree]$ and $[163\degree, 180\degree]$. 
These findings are in agreement with the results of \cite{EHTV2022}, who identified two promising MAD models with positive spin and low inclination that describe the supermassive black hole in our Galactic center.
Specifically, the parameters for the model associated with our GRMHD simulation read $(a, i)=(0.94, 30\degree)$ and align perfectly with the results of our analysis. 
These insights contribute to a more detailed understanding of accretion-driven magnetic activity and offer a basis for future observational comparisons, particularly with NIR and X-ray variability linked to episodic flares around supermassive black holes.
\section*{Data availability}
\url{https://zenodo.org/records/14843928}
\begin{acknowledgements}
The authors thank O. Porth and I. Contopoulos for fruitful discussions and comments on this work, A. Liberatori and F. Zanias for support in the calculations, and the anonymous referee for constructive comments. 
This work was supported by computational time granted from the National Infrastructures for Research and Technology S.A. (GRNET S.A.) in the National HPC facility - ARIS - under project ID 16033.  
AL acknowledges financial support by the State Scholarships
Foundation (IKY) scholarship program from the proceeds of the “Nic. D.
Chrysovergis” bequest.
\end{acknowledgements}
\nocite{*}
\bibliographystyle{aa} 
%
\bibliography{MAD_flux_eruption_events_to_describe_SgrA_flares}

\begin{thebibliography}{65}
\expandafter\ifx\csname natexlab\endcsname\relax\def\natexlab#1{#1}\fi

\bibitem[{{Aimar} {et~al.}(2023){Aimar}, {Dmytriiev}, {Vincent}, {El Mellah},
  {Paumard}, {Perrin}, \& {Zech}}]{Aimar_2023}
{Aimar}, N., {Dmytriiev}, A., {Vincent}, F.~H., {et~al.} 2023, \aap, 672, A62

\bibitem[{{Antonopoulou} \& {Nathanail}(2024)}]{Antonopoulou_2024}
{Antonopoulou} \& {Nathanail}. 2024, A\&A, 690, A240

\bibitem[{{Baganoff} {et~al.}(2001){Baganoff}, {Bautz}, {Brandt}, {Chartas},
  {Feigelson}, {Garmire}, {Maeda}, {Morris}, {Ricker}, {Townsley}, \&
  {Walter}}]{baganoff2001}
{Baganoff}, F.~K., {Bautz}, M.~W., {Brandt}, W.~N., {et~al.} 2001, \nat, 413,
  45

\bibitem[{{Ball, D., Özel, F., et al}(2021)}]{Ball_2021}
{Ball, D., Özel, F., et al}. 2021, The Astrophysical Journal, 917, 8

\bibitem[{{Bisnovatyi-Kogan} \& {Ruzmaikin}(1974)}]{Bisnovatyi_Ruzmaikin_1974}
{Bisnovatyi-Kogan}, G.~S. \& {Ruzmaikin}, A.~A. 1974, \apss, 28, 45

\bibitem[{{Bower} {et~al.}(2005){Bower}, {Falcke}, {Wright}, \&
  {Backer}}]{Bower_2005}
{Bower}, G.~C., {Falcke}, H., {Wright}, M.~C., \& {Backer}, D.~C. 2005, \apjl,
  618, L29

\bibitem[{{Chatterjee} {et~al.}(2021){Chatterjee}, {Markoff}, {Neilsen},
  {Younsi}, {Witzel}, {Tchekhovskoy}, {Yoon}, {Ingram}, {van der Klis},
  {Boyce}, {Do}, {Haggard}, \& {Nowak}}]{Chatterjee_2021}
{Chatterjee}, K., {Markoff}, S., {Neilsen}, J., {et~al.} 2021, \mnras, 507,
  5281

\bibitem[{Cruz-Osorio {et~al.}(2022)Cruz-Osorio, Fromm, Mizuno, Nathanail,
  Younsi, Porth, Davelaar, Falcke, Kramer, \& Rezzolla}]{CruzOsorio_2022}
Cruz-Osorio, A., Fromm, C., Mizuno, Y., {et~al.} 2022, Nature Astronomy, 6

\bibitem[{{Dexter, J., Tchekhovskoy, A, et al.}(2020)}]{Dexter_2020}
{Dexter, J., Tchekhovskoy, A, et al.} 2020, \mnras, 497, 4999

\bibitem[{{Dimitropoulos, I., et al.}(2024)}]{Dimitropoulos_2024}
{Dimitropoulos, I., et al.} 2024, submitted to A\&A, arXiv:2407.14312

\bibitem[{{Event Horizon Telescope Collaboration et al.}(2019)}]{EHT_M87_V}
{Event Horizon Telescope Collaboration et al.} 2019, ApJL, 875, L5

\bibitem[{{Event Horizon Telescope Collaboration et al.}(2022)}]{EHTV2022}
{Event Horizon Telescope Collaboration et al.} 2022, ApJL, 930, L16

\bibitem[{{Event Horizon Telescope Collaboration et al.}(2024)}]{EHTVIII2024}
{Event Horizon Telescope Collaboration et al.} 2024, \apjl, 964, L26

\bibitem[{Fermo {et~al.}(2010)Fermo, Drake, \& Swisdak}]{Fermo_2010}
Fermo, R.~L., Drake, J.~F., \& Swisdak, M. 2010, Physics of Plasmas, 17, 010702

\bibitem[{{Fishbone} \& {Moncrief}(1976)}]{Fishborne_Moncrief_1976}
{Fishbone}, L.~G. \& {Moncrief}, V. 1976, \apj, 207, 962

\bibitem[{{Fromm, C. M., Cruz-Osorio, A., et al.}(2022)}]{Fromm_2022}
{Fromm, C. M., Cruz-Osorio, A., et al.} 2022, A\&A, 660, A107

\bibitem[{{Genzel} {et~al.}(2003){Genzel}, {Sch{\"o}del}, {Ott}, {Eckart},
  {Alexander}, {Lacombe}, {Rouan}, \& {Aschenbach}}]{genzel2003}
{Genzel}, R., {Sch{\"o}del}, R., {Ott}, T., {et~al.} 2003, \nat, 425, 934

\bibitem[{{Ghez, A. M., et al.}(2004)}]{Ghez_2004}
{Ghez, A. M., et al.} 2004, The Astrophysical Journal, 601, L159

\bibitem[{Ghisellini(2013)}]{Ghisellini_2013}
Ghisellini, G. 2013, Radiative Processes in High Energy Astrophysics (Springer)

\bibitem[{{GRAVITY Collaboration et al.}(2018)}]{GRAVITY_2018}
{GRAVITY Collaboration et al.} 2018, A\&A, 618, L10

\bibitem[{{GRAVITY Collaboration et
  al.}(2020{\natexlab{a}})}]{GRAVITY_2020_polarimetry}
{GRAVITY Collaboration et al.} 2020{\natexlab{a}}, A\&A, 643, A56

\bibitem[{{GRAVITY Collaboration et al.}(2020{\natexlab{b}})}]{GRAVITY_2020}
{GRAVITY Collaboration et al.} 2020{\natexlab{b}}, \aap, 635, A143

\bibitem[{{GRAVITY Collaboration et al.}(2023)}]{GRAVITY_2023}
{GRAVITY Collaboration et al.} 2023, \aap, 677, L10

\bibitem[{{Huang} {et~al.}(2024){Huang}, {Zhang}, {Guo}, \&
  {Chen}}]{Huang_2024}
{Huang}, J., {Zhang}, Z., {Guo}, M., \& {Chen}, B. 2024, \prd, 109, 124062

\bibitem[{Huang \& Bhattacharjee(2012)}]{Huang_2012}
Huang, Y.-M. \& Bhattacharjee, A. 2012, Phys. Rev. Lett., 109, 265002

\bibitem[{Igumenshchev(2008)}]{Igumenshchev_2008}
Igumenshchev, I.~V. 2008, The Astrophysical Journal, 677, 317

\bibitem[{{Igumenshchev, I. V., et al}(2003)}]{Igumenshchev_2003}
{Igumenshchev, I. V., et al}. 2003, The Astrophysical Journal, 592, 1042

\bibitem[{{Kagan, D.,Sironi, L., et al.}(2015)}]{Kagan_2015}
{Kagan, D.,Sironi, L., et al.} 2015, \ssr, 191, 545

\bibitem[{{Kocherlakota, P., Rezzolla, L., et al.}(2024)}]{kocherlakota_2024}
{Kocherlakota, P., Rezzolla, L., et al.} 2024, \mnras, 531, 3606

\bibitem[{{Lin} {et~al.}(2023){Lin}, {Li}, \& {Yuan}}]{Lin_2023}
{Lin}, X., {Li}, Y.-P., \& {Yuan}, F. 2023, \mnras, 520, 1271

\bibitem[{{Lin} \& {Yuan}(2024)}]{Lin_2024}
{Lin}, X. \& {Yuan}, F. 2024, \mnras, 531, 3136

\bibitem[{{Loureiro, N. F., et al.}(2007)}]{Loureiro_2007}
{Loureiro, N. F., et al.} 2007, Physics of Plasmas, 14, 100703

\bibitem[{{Loureiro, N. F., Samtaney, R., et al.}(2012)}]{Loureiro_2012}
{Loureiro, N. F., Samtaney, R., et al.} 2012, Physics of Plasmas, 19, 042303

\bibitem[{{Marrone} {et~al.}(2007){Marrone}, {Moran}, {Zhao}, \&
  {Rao}}]{Marrone_2007}
{Marrone}, D.~P., {Moran}, J.~M., {Zhao}, J.-H., \& {Rao}, R. 2007, \apjl, 654,
  L57

\bibitem[{Matsumoto {et~al.}(2020)Matsumoto, Chan, \& Piran}]{Matsumoto_2020}
Matsumoto, T., Chan, C.-H., \& Piran, T. 2020, \mnras, 497, 2385

\bibitem[{McKinney \& Gammie(2004)}]{McKinney_2004}
McKinney, J.~C. \& Gammie, C.~F. 2004, The Astrophysical Journal, 611, 977

\bibitem[{{Moscibrodzka, M., Proga, D., et al}(2007)}]{Moscibr_2007}
{Moscibrodzka, M., Proga, D., et al}. 2007, \aap, 474, 1

\bibitem[{{Najafi-Ziyazi, M. , Davelaar, J., et al}(2024)}]{mahdi_2023}
{Najafi-Ziyazi, M. , Davelaar, J., et al}. 2024, \mnras, 531, 3961

\bibitem[{{Narayan} {et~al.}(2003){Narayan}, {Igumenshchev}, \&
  {Abramowicz}}]{Narayan_2003}
{Narayan}, R., {Igumenshchev}, I.~V., \& {Abramowicz}, M.~A. 2003, \pasj, 55,
  L69

\bibitem[{{Narayan, R., Chael, A., et al.}(2022)}]{narayan2022}
{Narayan, R., Chael, A., et al.} 2022, \mnras, 511, 3795

\bibitem[{{Nathanail} {et~al.}(2022){Nathanail}, {Dhang}, \&
  {Fromm}}]{Nathanail_2022b}
{Nathanail}, A., {Dhang}, P., \& {Fromm}, C.~M. 2022, \mnras, 513, 5204

\bibitem[{{Nathanail} {et~al.}(2020){Nathanail}, {Fromm}, {Porth}, {Olivares},
  {Younsi}, {Mizuno}, \& {Rezzolla}}]{Nathanail_2020}
{Nathanail}, A., {Fromm}, C.~M., {Porth}, O., {et~al.} 2020, \mnras, 495, 1549

\bibitem[{{Nathanail, A., Mpisketzis, V., et al.}(2022)}]{Nathanail_2022}
{Nathanail, A., Mpisketzis, V., et al.} 2022, \mnras, 513, 4267

\bibitem[{{Olivares} {et~al.}(2019){Olivares}, {Porth}, {Davelaar}, {Most},
  {Fromm}, {Mizuno}, {Younsi}, \& {Rezzolla}}]{Olivares_2019}
{Olivares}, H., {Porth}, O., {Davelaar}, J., {et~al.} 2019, \aap, 629, A61

\bibitem[{{Pandya, A., Zhang, et al.}(2016)}]{Pandya_2016}
{Pandya, A., Zhang, et al.} 2016, The Astrophysical Journal, 822, 34

\bibitem[{Papadopoulos \& Contopoulos(2018)}]{Papadopoulos_2018}
Papadopoulos, D.~B. \& Contopoulos, I. 2018, \mnras, 483, 2325

\bibitem[{{Pasham, D. R., et al.}(2024)}]{Pasham_2024}
{Pasham, D. R., et al.} 2024, Science Advances, 10, eadj8898

\bibitem[{Porth {et~al.}(2021)Porth, Mizuno, Younsi, \& Fromm}]{Porth_2021}
Porth, O., Mizuno, Y., Younsi, Z., \& Fromm, C.~M. 2021, \mnras, 502, 2023

\bibitem[{{Porth, O., et al.}(2017)}]{Porth_2017}
{Porth, O., et al.} 2017, Computational Astrophysics and Cosmology, 4, 1

\bibitem[{Ressler {et~al.}(2019)Ressler, Quataert, \& Stone}]{Ressler_2019}
Ressler, S.~M., Quataert, E., \& Stone, J.~M. 2019, \mnras, 492, 3272

\bibitem[{{Ripperda, B., Liska, M., et al.}(2022)}]{Ripperda_2022}
{Ripperda, B., Liska, M., et al.} 2022, The Astrophysical Journal Letters, 924,
  L32

\bibitem[{{Rybicki} \& {Lightman}(1986)}]{Rybicki_1986}
{Rybicki}, G.~B. \& {Lightman}, A.~P. 1986, {Radiative Processes in
  Astrophysics}

\bibitem[{{Sukov{\'a}} {et~al.}(2021){Sukov{\'a}}, {Zaja{\v{c}}ek}, {Witzany},
  \& {Karas}}]{Sukova_2021}
{Sukov{\'a}}, P., {Zaja{\v{c}}ek}, M., {Witzany}, V., \& {Karas}, V. 2021,
  \apj, 917, 43

\bibitem[{Takamoto(2013)}]{Takamoto_2013}
Takamoto, M. 2013, The Astrophysical Journal, 775, 50

\bibitem[{{Tchekhovskoy, A., et al.}(2011)}]{Tchekhovskoy_2011}
{Tchekhovskoy, A., et al.} 2011, Monthly Notices of the Royal Astronomical
  Society: Letters, 418, L79

\bibitem[{{Tursunov} {et~al.}(2020){Tursunov}, {Zaja{\v{c}}ek}, {Eckart},
  {Kolo{\v{s}}}, {Britzen}, {Stuchl{\'\i}k}, {Czerny}, \&
  {Karas}}]{Tursunov_2020}
{Tursunov}, A., {Zaja{\v{c}}ek}, M., {Eckart}, A., {et~al.} 2020, \apj, 897, 99

\bibitem[{{Uzdensky, D. A., et al.}(2010)}]{Uzdensky_2010}
{Uzdensky, D. A., et al.} 2010, Phys. Rev. Lett., 105, 235002

\bibitem[{{Van der Laan}(1966)}]{VanderLaan1966}
{Van der Laan}, H. 1966, \nat, 211, 1131

\bibitem[{{Vincent, F.~H., Wielgus, M., et al.}(2024)}]{vincent_2023}
{Vincent, F.~H., Wielgus, M., et al.} 2024, \aap, 684, A194

\bibitem[{{Vos} {et~al.}(2022){Vos}, {Mo{\'s}cibrodzka}, \&
  {Wielgus}}]{Vos_2022}
{Vos}, J., {Mo{\'s}cibrodzka}, M.~A., \& {Wielgus}, M. 2022, \aap, 668, A185

\bibitem[{Wang {et~al.}(2013)Wang, Nowak, Markoff, Baganoff, Nayakshin, Yuan,
  Cuadra, Davis, Dexter, Fabian, Grosso, Haggard, Houck, Ji, Li, Neilsen,
  Porquet, Ripple, \& Shcherbakov}]{Wang_2013}
Wang, Q.~D., Nowak, M.~A., Markoff, S.~B., {et~al.} 2013, Science, 341,
  981–983

\bibitem[{{Wong} {et~al.}(2021){Wong}, {Du}, {Prather}, \& {Gammie}}]{wong2021}
{Wong}, G.~N., {Du}, Y., {Prather}, B.~S., \& {Gammie}, C.~F. 2021, \apj, 914,
  55

\bibitem[{{Yfantis, A. I., Mo{\'s}cibrodzka, M. A., et
  al.}(2024)}]{yfantis_2023}
{Yfantis, A. I., Mo{\'s}cibrodzka, M. A., et al.} 2024, \aap, 685, A142

\bibitem[{Yuan {et~al.}(2009)Yuan, Lin, Wu, \& Ho}]{Yuan_2009}
Yuan, F., Lin, J., Wu, K., \& Ho, L.~C. 2009, \mnras, 395, 2183

\bibitem[{{Zhang} {et~al.}(2024){Zhang}, {B{\'e}gu{\'e}}, {Pe'er}, \&
  {Zhang}}]{zhang2024}
{Zhang}, G.~Q., {B{\'e}gu{\'e}}, D., {Pe'er}, A., \& {Zhang}, B.~B. 2024, \apj,
  962, 135

\end{thebibliography}

\begin{appendix} 
\section{Numerical setup} \label{Append:Numerical_Setup}
This study investigates a MAD simulation, motivated by the best-bet models of \cite{EHTV2022}.
The simulation is performed in three spatial dimensions with BHAC \citep{Porth_2017,Olivares_2019}\footnote{\, https://www.bhac.science}, using modified Kerr-Schild coordinates \citep{McKinney_2004} and 2–3 levels of adaptive mesh refinement.
We employed units where $c=G=1$, which corresponds to a length unit equal to the black hole mass $r_g=M$, and a base resolution of $384 \times 192 \times 192$ grid cells in the $r$, $\theta$, and $\phi$ directions, respectively. 
\begin{table*}     
\begin{center}                       
\begin{tabular}{c c c c c c c c}  
\hline\hline 
Model & a & Resolution & Duration $(M)$ & $\hat{\gamma}$ & $r_{in} (M)$ & $r_{max} (M)$ & $r_{out} (M)$  \\
\hline
MAD & 0.94 & $384\times192\times192$ & 100,000 & 4/3 & 20 & 40 & 2,500 \\ 
\hline   
\end{tabular}
\end{center}
\caption{GRMHD simulation parameters. Model specifies the accretion disk model, $a$ is the dimensionless black hole spin, Resolution denotes the number of grid zones along each direction as $N_r\times N_{\theta}\times N_{\phi}$, Duration denotes the total duration of the simulation, $\gamma$ is the fluid element adiabatic index, $r_{in}$ and $r_{max}$ are the inner and pressure maximum radii of the FM torus, and $r_{out}$ is the outer radial boundary of the simulation domain.}       
\label{Table:GRMHD_Setup} 
\end{table*}
The initial conditions consist of a \cite{Fishborne_Moncrief_1976} torus in hydrodynamic equilibrium, with a constant specific angular momentum of $l=6.76$, orbiting a Kerr black hole with a dimensionless spin parameter of $a=0.94$.
The inner edge of the disk lies at $r_{in}= 20 M$, the density maximum is located at $r_{max} = 40 M$, and the equation of state is ideal, with an adiabatic index of $\hat{\gamma}= 4/3$.
Furthermore, the magnetic field is initialized as a nested loop described by the vector potential
\begin{equation}
    \label{vector_potential}
    A_{\phi} = max \left( \left( \dfrac{\rho}{\rho_{max}} \, \left(\dfrac{r}{r_{in}}\right)^3 \, \sin^3\theta\,\exp{\left(\dfrac{-r}{400}\right)} \right) - 0.2, 0 \right) \,,
\end{equation}
where $\rho_{max}$ is the maximum rest-mass density in the torus 
\citep{EHT_M87_V,wong2021,CruzOsorio_2022,Fromm_2022,narayan2022,EHTV2022,zhang2024}.
The field strength is set such that $2P_{max}/B^2_{max}=100$, where the global maxima of pressure $P_{max}$ and magnetic field strength $B^2_{max}$ do not necessarily coincide.
With this choice of initial magnetic field geometry and strength, the simulation progresses according to the MAD regime (see Appendix \ref{Append:MAD_Time_Series}).
MAD models are characterized by high values of magnetic flux at the event horizon $\phi_{BH} \sim \phi_{crit} \sim 50$ and demonstrate a quasiperiodic cycle of flux eruption events.
In particular, magnetic flux accretes onto the black hole until it reaches the saturation value $\phi_{BH}\geq\phi_{crit}$.
Thereafter, accretion of additional flux leads to flux expulsion events, during which $\phi_{BH}$ rapidly decreases. 
In contrast, SANE models are characterized by low values of magnetic flux $\phi_{BH}\ll\phi_{crit}$ and do not demonstrate the characteristic flux eruption events of the MAD accretion state \citep{CruzOsorio_2022,Fromm_2022}.
As a result, this study does not investigate SANE accretion disk simulations.
\par
Radiative transfer calculations are performed in Boyer-Lindquist coordinates with the Black Hole Imaging code, first presented in \cite{Antonopoulou_2024}.
A Kerr black hole, with dimensionless spin parameter $a=0.94$, rests at the center of the coordinate system, and a grid of $600\times600$ photons is initialized at a distance of $500M$ from the center.
Unless clearly stated otherwise, observer inclination is face-on $-i=5\degree$ to eliminate issues on the polar axis$-$ throughout our calculations.
In the following calculations, hot spot emission is modeled by a power-law electron distribution \citep{Pandya_2016}, while synchrotron absorption is neglected as a first-order approximation. 
\section{MAD accretion state} \label{Append:MAD_Time_Series}
\begin{figure}[t]
\centering
\includegraphics[width=0.49\textwidth]{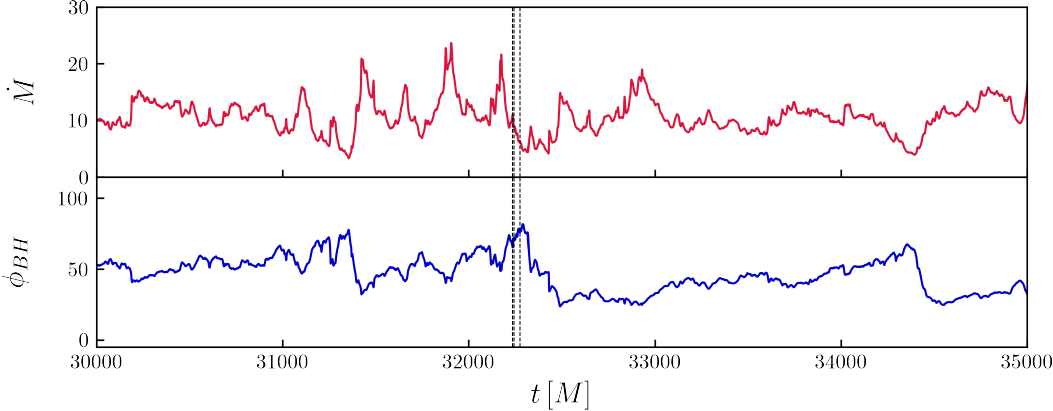}
\caption{Time series of the mass accretion rate (top panel) and the normalized magnetic flux (bottom panel) throughout the simulation. The dashed vertical lines indicate the times displayed in Fig. \ref{Fig:Flux_eruption}.}
\label{Fig:Mdot_Phi}
\end{figure}

We defined the mass accretion rate and magnetic flux over the surface of the black hole's horizon, as follows:
\begin{align}
    \dot{M} &\equiv -\iint \rho u^r \sqrt{-g}\,d\theta\,d\phi \,, \\
    \Phi_{BH} &\equiv \dfrac{1}{2}\iint \,\abs{B^r}\sqrt{-g}\,d\theta\,d\phi \,.
\end{align}
where, $\rho, u^r, g,$ and $B^r$ are the mass density, radial four-velocity, metric determinant, and radial component of the magnetic field as seen by the local fiducial observers, respectively.
The dimensionless normalized magnetic flux at the horizon can then be defined as
\begin{equation}
    \phi_{BH}\equiv \sqrt{4\pi}\:\dfrac{\Phi_{BH}}{\sqrt{\dot{M}}} \,.
\end{equation}
Figure \ref{Fig:Mdot_Phi} shows the time series of the mass accretion rate and the dimensionless normalized magnetic flux during the selected flux eruption event.
By time $30.000M$, the simulation has already reached a quasi-stationary MAD state, and quasiperiodic dips in the normalized magnetic flux can be identified throughout the bottom part of the figure. 
These dips correspond to flux eruption events in the accretion disk, where the magnetic flux is expelled and matter accretion is significantly halted. 
They occur when the parameter $\phi_{BH}$ reaches its saturation limit ($\phi_{BH}\sim 50$; \citealt{Tchekhovskoy_2011}).  
The dashed vertical lines indicate the snapshots of the flaring event illustrated in Fig. \ref{Fig:Flux_eruption}.
\section{Modeling MAD flares} \label{Append:Modeling_MAD_flares}
Our basic assumptions in this work are the following: during magnetic reconnection, the released magnetic energy generates energetic particles, forming hot spots on the accretion disk’s equatorial plane, spanning a fraction of the gravitational radius. 
These hot spots are then ejected from the disk plane and travel along highly magnetized flux tubes at a fraction of the speed of light.
It is clear from the evolution of the simulation that as the flux bundle moves counterclockwise with the disk’s accretion flow, it experiences bending in the opposite direction due to rotation. 
Consequently, hot spots trapped within these flux bundles are observed to move in a clockwise manner, opposite to the general flow of the accretion disk.
The focus of Sect. \ref{Sec:Methodology} was set on identifying a promising flux eruption event candidate, capable of replicating the flaring observations in the vicinity of SgrA*, as well as tracking the evolution of both the magnetic reconnection site on the disk's equator and the magnetic field lines comprising the energetic flux tube. 
The following paragraphs outline the modeling procedure adopted for the selected flux eruption event, illustrated in Fig. \ref{Fig:Flux_eruption}. \par
This model considers spherical hot spots with a constant radius of $r_{\rm spot}=0.5M$ and a radiative sphere of radius $r_{\rm rad}=1M$.
In particular, the interior of the hot spot is considered to be optically thick; therefore, the total nonthermal synchrotron emission originates from the radiative sphere surrounding it.
Although the shape of the hot spot is certainly subjected to continuous alterations throughout the course of its outward trajectory (due to shearing, for instance), the precise mechanism behind this process strongly depends on the morphology of the highly magnetized surrounding plasma and has yet to be investigated in a concise manner.
Due to a lack of a concise analytic prescription, a constant spherical hot spot shape is maintained throughout the radiative transfer calculations, with an effective emission radius of $1M$.
\par
The hot spots are in principle generated at the equatorial plane of the disk, fueled by the magnetic reconnection process, and ejected outward along the magnetized flux tube.
As a first approximation, we assumed that the generated hot spot gains a large amount of energy, enough to traverse along the flux tube with a relativistic velocity of a fraction of the speed of light.
We further assumed that the energetic hot spot effectively maintains a constant ejection velocity throughout its trajectory.
In the present analysis, the ejection velocity ranges from $0.5$ up to $0.8$ times the speed of light for individual models. 
\par
Another important parameter to be taken into account is the motion of the active magnetic reconnection site on the equatorial plane of the disk.
As discussed in Sect. \ref{Sec:Methodology}, the active region exhibits two distinct motion patterns throughout the flaring event.
Firstly, one must consider the outward motion, toward larger orbital radii, by carefully tracking the position of the flux tube's foot-point on the equator of the accretion disk.
For the flaring event illustrated in Fig. \ref{Fig:Flux_eruption}, this corresponds to an average radial velocity of $0.07c$.
Secondly, one must account for the orbital motion of the active region, as a result of the accretion disk's rotation. 
However, the selection of magnetic field lines and foot-points on the equatorial plane introduces some deviations in the flux tube’s rotation.
In particular, the flux tube continuously changes shape during the flux eruption event, rendering the measurement of the precise orbital velocity for each foot-point a difficult task \citep{Porth_2021}.
This analysis assumes a solid-body rotation for the flux tube and investigates sub-Keplerian orbital frequencies ranging from $0.1u_K-0.4u_K$ for individual models, based on the simulation.
\section{Flare energetics} \label{Append:Flare_Energetics}
The flare models presented in this study are in good agreement with the kinematics of the prominent NIR flare observed on July 22, 2018, as illustrated in Fig. \ref{Fig:Flare_Observations}.
This section focuses on flare energetics, providing an analytical estimate of the flux emitted at $2.2\mu m$ (K-band) based on our GRMHD simulation parameters, and a brief discussion of the flare cessation mechanism.
\par
As described in Appendix \ref{Append:Numerical_Setup}, all calculations throughout the paper consider a nonthermal electron population with density
\begin{equation}
    \label{eq:power-law}
    N(\gamma) = n_e\,\gamma^{-p}, \quad\gamma_{min}<\gamma<\gamma_{max} \,,
\end{equation}
where $n_e$ is the electron number density, $p$ is the slope of the power-law distribution, and $\gamma_{min}$, $\gamma_{max}$ are the minimum and maximum Lorentz factor for the electrons, respectively.
The pitch angle averaged synchrotron emissivity is thus given by \citep{Rybicki_1986,Ghisellini_2013}
\begin{equation}
    \label{eq:synchr_jv}
    j_s(v) = \dfrac{3c\sigma_Tn_e U_B}{16\pi \sqrt{\pi}v_L}\,\left(\dfrac{v}{v_L}\right)^{-\frac{p-1}{2}} \,f_j(p)\,,
\end{equation}
where $U_B=B^2/8\pi$ is the energy density of the magnetic field.
In the expression above, function $f_j(p)$ includes the products of the $\Gamma-$functions and can be approximated as
\begin{equation}
    \label{eq:fj_p}
    f_j(p) \sim 3^\frac{p}{2}\,\left(\dfrac{2.25}{p^{2.2}} + 0.105 \right) \,.
\end{equation}
Assuming optically thin emission, the flux density from a spherical hot spot is given by \citep{EHTV2022}
\begin{equation}
    \label{eq:flux_estimate}
    F_v = \dfrac{4}{3}\pi r_{spot}^3\,j_s\,D^{-2}\,10^{23}\;Jy\,,
\end{equation}
where $r_{spot}=1r_g$ is the hot spot radius for our flare models and $D$ is the distance from SgrA*.
We assumed an average accretion rate of $\dot{M}\sim 10^{-9}\,M_{\odot}\,yr^{-1}$ for SgrA* \citep{Bower_2005,Marrone_2007,Wang_2013,EHTV2022}, corresponding to an electron number density of $n_e\simeq10^6\,cm^{-3}$ and a magnetic field strength of $B\simeq87\,G$ for our MAD simulation.
For a power-law index $p=3.6$, the emitted flux at $2.2\mu m$ is equal to $F_v\simeq 7.85\,mJy \simeq 0.52\,F(S2)$, where $F(S2)\simeq15mJy$ is the K-band flux density of the star S2.  
This analytical estimate is in good agreement with the observed peak flux of the bright July 22 flare, which reached a value of $F_{peak}\simeq0.6\,F(S2)$ \citep{GRAVITY_2018}. 
\par
Although our calculations assume a constant spherical hot spot shape, the mechanism responsible for the flare cessation is adiabatic cooling, dating back to the \cite{VanderLaan1966} model.
In particular, as the hot spot expands in size, the relativistic gas cools adiabatically, the magnetic field strength decreases, and the synchrotron emission shifts to lower frequencies.
The adiabatic cooling timescale is thus proportional to
\begin{equation}
    t_{ad} \propto \dfrac{R-R_0}{\dot{R}} \,.
\end{equation}
Assuming that the ejected hot spot doubles in size during the flaring event and considering a constant expansion rate of $\dot{R}\sim 0.005$, the adiabatic cooling timescale is proportional to the duration of the observed flares, $t_{ad}\propto 200\,M$.
While hot spot expansion is not considered in this work, due to the lack of a precise analytic prescription (a more detailed discussion is given in Appendix \ref{Append:Modeling_MAD_flares}), our analytical estimate shows that the adiabatic cooling mechanism is able to account for the characteristic duration of the observed flaring events in our Galactic center of $\sim1\,$hour. 
\section{Reproducing the first data points}\label{Append:GRAVITY_first_observation}
The \cite{GRAVITY_2018} has employed four independent codes to fit the signal of SgrA*.
They report that all four codes agree on the main features and results of the observed flares, despite their individual differences (fitting approach, number of free parameters, etc.).
The flare positions investigated throughout the paper represent the average of the Waisberg and Pfuhl analyses, in agreement with previous studies.
This methodology effectively incorporates the individual differences between the unique codes and reflects the systematic uncertainty of the GRAVITY observations. 
For instance, the Waisberg analysis of the July 22 flare results in a significantly less elongated loop in the y-direction than both the Pfuhl analysis and the averaged data (Fig. B.1. of \citealt{GRAVITY_2018}).
As a result, considering the Waisberg flare positions would allow configurations that produce less extended orbits to fit the GRAVITY observations.
However, since the key flare features -- such as the kinematics of the first data points -- are present in both analyses, we do not expect this choice to significantly affect our results.
\par
The focus of this section is set on reproducing the first data points of the July 22 flare.
While several models have attempted to replicate the prominent flare of July 22, none has successfully captured the assumed kinematics of the first data point (notice the dark red point at $t=1.03$min).
Although the flare demonstrates a clear clockwise motion in the sky plane, the first flaring position exhibits a sudden push in the opposite direction (notice the bright red point at $t=3.59$min), before continuing along the observed clockwise pattern.
\par
\begin{figure}[t]
\includegraphics[width=0.45\textwidth]{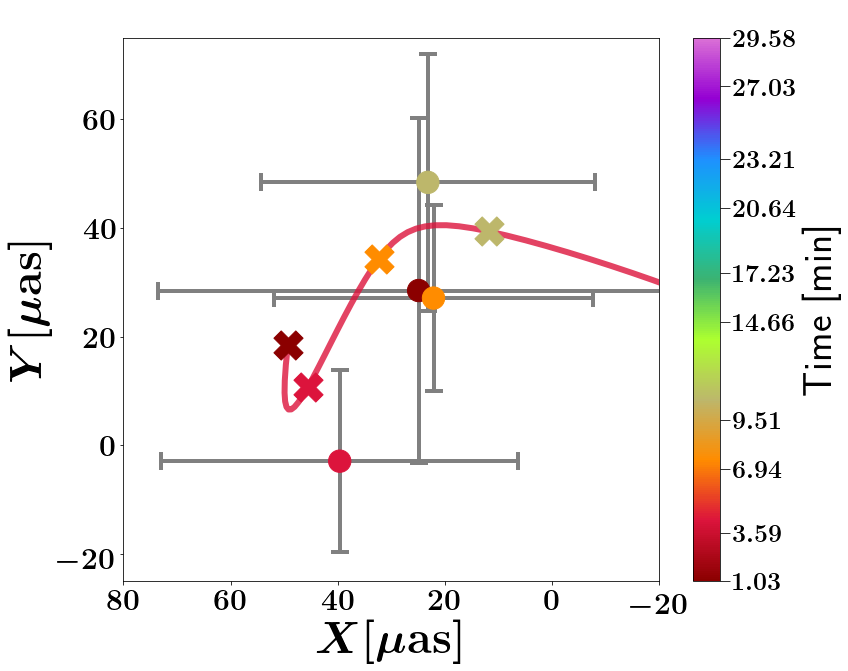} 
\includegraphics[width=0.45\textwidth]{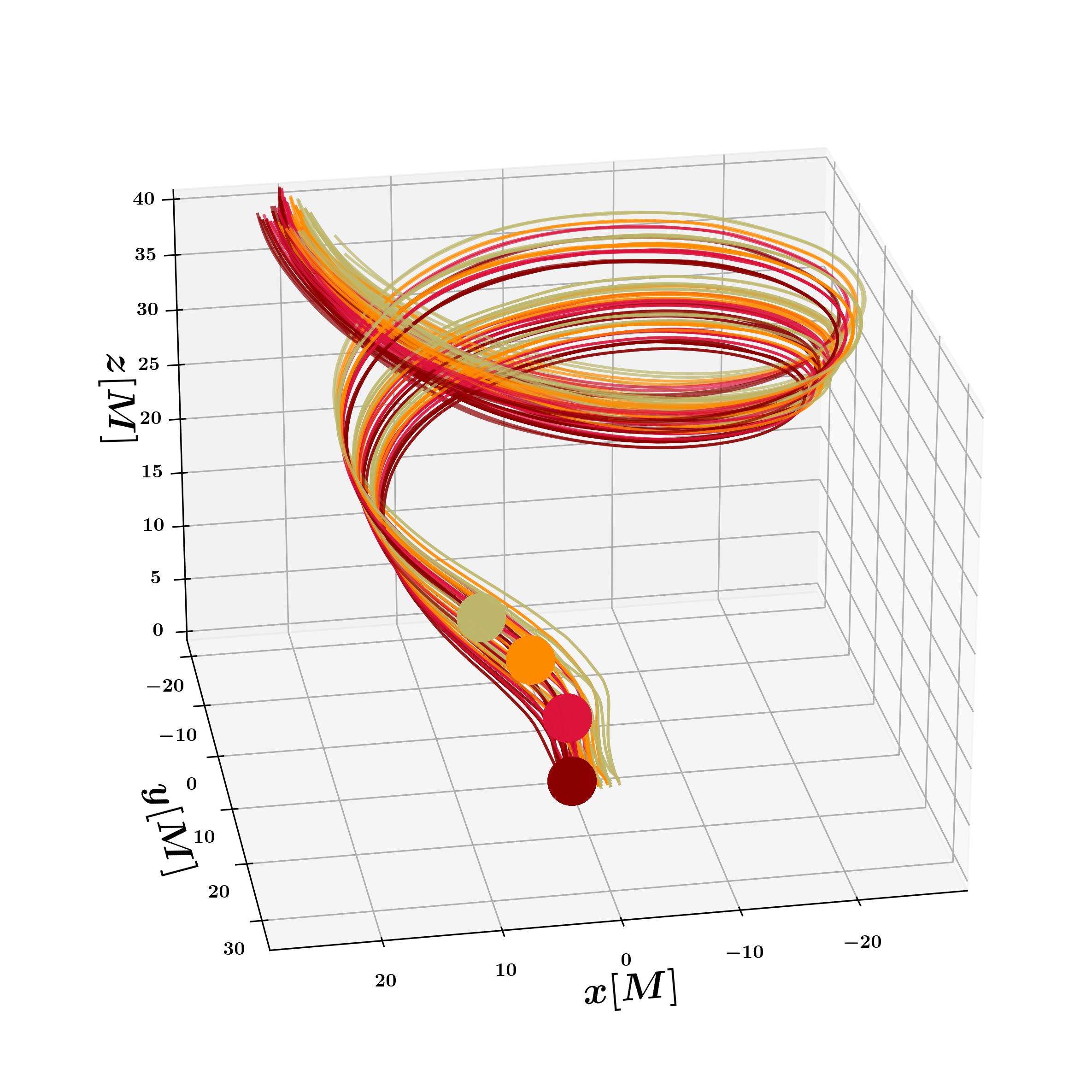}
\caption{One of the flare models illustrated in Fig. \ref{Fig:Flare_Observations} capable of capturing the kinematics of the first data points. \textit{Top panel}: Close-up view of the selected flare model overlapped with the first four observed flaring positions. \textit{Bottom panel}: Three-dimensional visualization of the corresponding flux tube. The hot spot positions for the first four data points are color-coordinated with the associated points in the top panel.}
\label{Fig:First_Observation}
\end{figure}
Figure \ref{Fig:First_Observation} depicts a close-up view of the first four observations for the July 22 flare, overlapped with one of the hot spot models illustrated in Fig. \ref{Fig:Flare_Observations} (left panel), as well as a three-dimensional visualization of the corresponding flux tube (right panel).
This flare model corresponds to an orbital velocity of $0.4u_K$ for the foot-point on the disk, and an ejection velocity of $0.8c$ for the hot spot.
The position of the hot spot along the flux tube for the first four GRAVITY observations is color-coordinated in both panels of Fig. \ref{Fig:First_Observation}.
We note that the shape of the magnetized flux tube demonstrates a continuous clockwise loop and an upward trend, as well as a distinct counterclockwise motion due to the rotation of the accretion disk. 
Even though the hot spot is continuously gaining height, it appears to be passing very close to its initial position (notice the dark red and orange points) due to the projection of the flux tube's shape on the sky plane.
The resulting hot spot motion is in great agreement with the observed flaring positions at the beginning of the flaring event, as is evident in Fig. \ref{Fig:First_Observation}.
Consequently, our analysis naturally recovers the observed flaring behavior of \cite{GRAVITY_2018}.
\section{Constraining the inclination of SgrA*}\label{Append:inclination_SgrA}
\begin{figure*}
\sidecaption
\includegraphics[width=6cm]{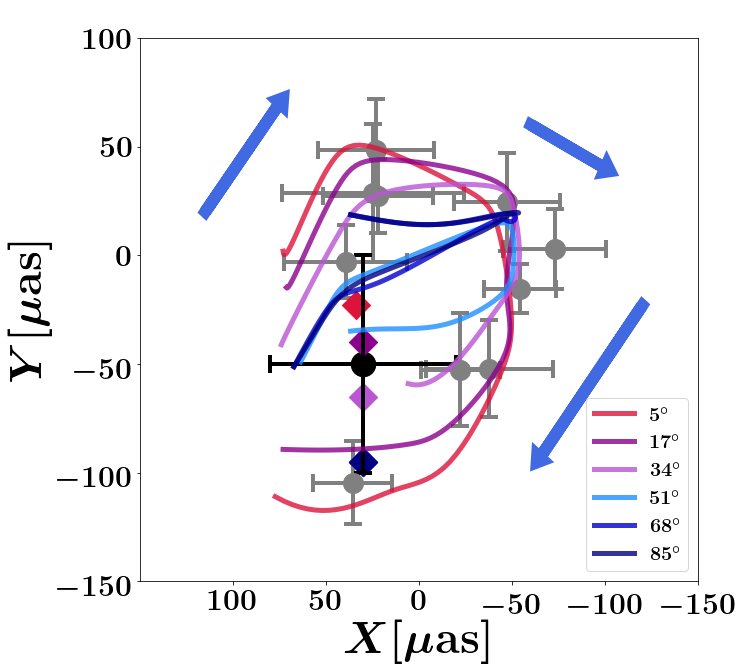}
\hspace{0.5cm}
\includegraphics[width=5.64cm]{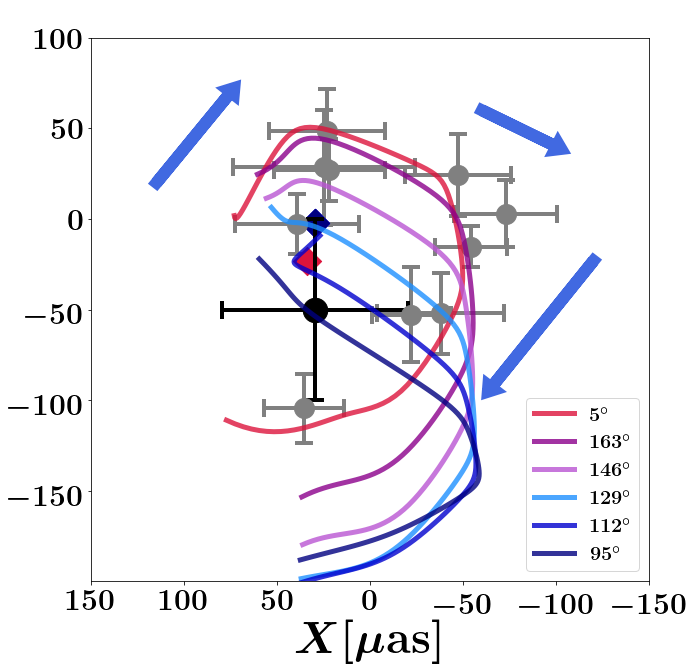} 
\caption{Best-fit flare model from Fig. \ref{Fig:Flare_Observations} for various inclinations (see the legends in the bottom right  of each panel) overlapped with the observations of July 22, 2018 (gray circles). The direction of motion is indicated by blue arrows. The black hole position for each inclination is color-coordinated with the corresponding hot spot trajectory and illustrated by a diamond. \textit{Left panel}: Inclinations between $[0\degree, 90\degree]$. \textit{Right panel}: Inclinations between $[90\degree, 180\degree]$.}
\label{Fig:Inclinations}
\end{figure*}
The trajectories illustrated in Fig. \ref{Fig:Flare_Observations} correspond to realistic flux tube configurations that are self-consistently generated throughout the flux eruption events observed in a MAD accretion disk simulation.
The following analysis investigates the image of the best fit hot spot trajectory (see the right panel of Fig. \ref{Fig:Flare_Observations}) for a distant observer at various inclinations.
\par
Figure \ref{Fig:Inclinations} illustrates the best fit hot spot model for an evenly spaced range of observation angles up to $85\degree$ (left panel), as well as their supplementary angles (right panel).
We note that the best fit trajectory of Fig. \ref{Fig:Flare_Observations} is depicted in red in both panels. 
It is evident that larger observation angles produce increasingly deformed hot spot orbits.
Specifically, observation angles between $[0\degree, 90\degree]$ result in oblate trajectories that are shrunk in the $y-$direction, whereas observation angles in the range $[90\degree, 180\degree]$ exhibit the opposite behavior, producing increasingly elongated orbits. 
The GRAVITY observations demonstrate a strong preference for face-on inclinations, as illustrated in Fig. \ref{Fig:Inclinations}.
Observing the flares at an edge-on inclination introduces large deformations in the shape of the orbit and pushes the hot spot trajectory beyond the bounds of the observed flaring emission in the Galactic center.
In particular, the present analysis constrains the inclination of SgrA* to the face-on range between $[0\degree, 34\degree]$ and $[163\degree, 180\degree]$.
On the other hand, hot spot trajectories observed at an inclination higher than $34\degree$ and lower than $163\degree$ are incompatible with the observed flaring behavior due to their large degree of deformation.
\section{Clockwise disk rotation} \label{Append:clockwise_flares}
Throughout the paper, we have investigated the quasiperiodic flux eruption events of a MAD accretion disk with a distinct counterclockwise rotation.
The generated flux tubes in this configuration demonstrate a clockwise pattern on the sky, as illustrated in Fig. \ref{Fig:First_Observation}.
Section \ref{Sec:Results} demonstrated that flare models with a small sub-Keplerian disk rotation and a relativistic hot spot velocity are in good agreement with the observed flaring behavior in the Galactic center.
On the contrary, if the accretion disk was rotating in a clockwise manner, the magnetized flux tubes would exhibit a counterclockwise pattern on the sky.
This section briefly investigates the scenario of clockwise accretion disk rotation.
\par
To consider a clockwise rotation in our MAD simulation, one has to observe the accretion disk from below ($\theta_{obs}=\pi$).
When viewed from below, the disk is indeed rotating in a clockwise manner with respect to the observer and the flux tubes demonstrate a distinct counterclockwise pattern on the sky.
We studied the evolution of the selected flux eruption event (see Fig. \ref{Fig:Flux_eruption}) and modeled the motion of hot spots along the generated flux tubes, following the formulation of Appendix \ref{Append:Modeling_MAD_flares}. 
Figure \ref{Fig:clockwise_flares} illustrates a selection of flare models (red lines) overlapped with the observations of July 22, 2018 (gray circles).
The depicted models correspond to various flux tube configurations during the selected flux eruption event, a hot spot ejection velocity equal to $0.1c$, and a clockwise disk rotation ranging from $0.8u_K$ to $0.9u_K$.
When the flux tube is rotating in a clockwise manner -in accordance with the observed flares-, the motion of the ejected hot spot along the counterclockwise shaped magnetic field lines is opposed to the clockwise evolution of the observed flare on the sky. 
Figure \ref{Fig:clockwise_flares} demonstrates that flare models with a nearly Keplerian disk rotation are able to balance out the ejected hot spot velocity and exhibit a distinct clockwise evolution that is in good agreement with the observed flares in the vicinity of SgrA*.
However, the rotation of the accretion disk in MAD simulations is well below the Keplerian limit and does not typically reach the values necessary to account for the flare models illustrated in Fig. \ref{Fig:clockwise_flares} \citep{Porth_2021}. 
\begin{figure}
\includegraphics[width=0.4\textwidth]{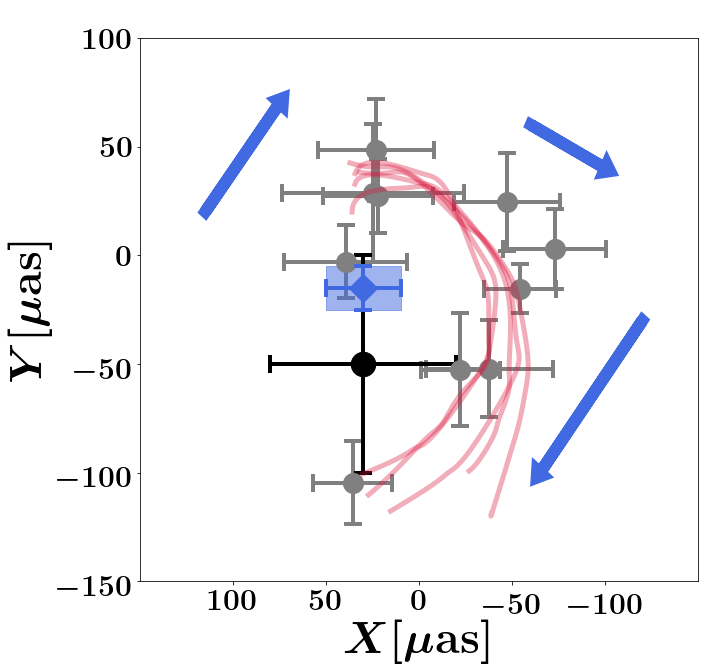} 
\caption{Flare models for a hot spot ejection velocity of $0.1c$ and foot-point orbital velocities between $0.8u_K$ and $0.9u_K$ (red lines) overlapped with the observations of July 22, 2018 (gray circles). The direction of motion is indicated by blue arrows. The position of SgrA* in the sky is denoted by a black cross, and the best-fit black hole position, which is slightly different for each model, by a blue rectangle.}
\label{Fig:clockwise_flares}
\end{figure}
\end{appendix}
\end{document}